\documentclass[twocolumn,showpacs,preprintnumbers]{revtex4}
\usepackage{amsmath,amssymb}
\usepackage{graphicx}
\usepackage{dcolumn}
\usepackage{bm}

\begin{document}

\title{Preparation of a Single Photon in Time-bin Entangled States via
Photon Parametric Interaction}
\author{N.Sisakyan}
\author{Yu.Malakyan}
\email{yumal@ipr.sci.am} \affiliation{Institute for Physical
Research, Armenian National Academy of Sciences, Ashtarak-2,
378410, Armenia }
\date{\today }

\begin{abstract}
A novel method for preparation of a single photon in
temporally-delocalized entangled modes is proposed and analyzed.
We show that two single-photon pulses propagating in a driven
nonabsorbing medium with different group velocities are temporally
split under parametric interaction into well-separated pulses. In
consequence, the single-photon "time-bin-entangled" states are
generated with a programmable entanglement easily controlled by
driving field intensity. The experimental study of nonclassical
features and nonlocality of generated states by means of balanced
homodyne tomography is discussed.
\end{abstract}

 \maketitle




\


Entanglement and nonlocal correlations, besides their fundamental
importance in the modern interpretation of quantum phenomena
\cite{ein, bell}, are the basic concepts for realization of
quantum information procedures \cite{ben}. The entanglement
between matter and light states is an essential element of quantum repeaters \cite%
{breig}, the intermediate memory nodes in quantum communication
network aimed at preventing the photon attenuation over long
distances. The two-photon entanglement is a crucial ingredient for
quantum cryptography \cite{ekert, tittel}, quantum teleportation
\cite{bouw, furus, marc}, and entanglement swapping \cite{zuk,
pan}, which have been successfully realized during the last decade
by utilizing two approaches, one based on continuous quadrature
variables and the other using the polarization variables of
quantized electromagnetic field \cite{braun}. An essential step
has been recently made in this direction by implementing robust
sources producing the pairs of photons which are entangled in
well-separated temporal modes (time-bins) \cite{brend}. It has
been shown \cite{brend, ried} that this type of entanglement, in
contrast to other ones, can be transferred over significant large
distances without appreciable losses that makes it much preferable
for long-distance applications. From the fundamental viewpoint, of
special interest is a single photon delocalized into two distinct
spatial \cite{knill} or temporal modes, for which case the
nonlocality of quantum correlations is directly evident from the
violation of Bell's inequality formulated for the two-mode Wigner function \cite%
{bonas}. This was verified experimentally by performing the
homodyne detection of delocalized single-photon Fock states and
reconstructing the corresponding Wigner function from homodyne
data \cite{lvov,babi, zava}. To date two approaches have been
developed for preparation of a single-photon in two distinct
temporal modes. In first one a time-bin qubit is created with the
help of linear optics by passing a short pulse through
Mach-Zehnder interferometer with different-length arms
\cite{brend}. The second approach is based on conditional
measurement on quantum system of entangled signal-idler pairs
generated via spontaneous parametric down conversion (SPDC) of
successive pump pulses in a nonlinear crystal, when a detection of
one idler photon tightly projects the signal field into a
single-photon state coherently delocalized over two temporal modes \cite%
{zava}.

In this paper we demonstrate a novel method for dynamical
preparation of time-bin qubit. The basic idea is to create a
parametric interaction between two single-photon pulses, which
propagate in a driven medium without absorption and with slow, but
different group velocities. Then, due to the cyclic parametric
conversion of the fields and the group delay, each pulse
experiences a temporal splitting into well-separated subpulses.
Moreover, since the process is completely coherent, at the output
of the medium the time-delocalized single-photon states are
formed. A remarkable feature of our scheme is the ability to
produce two and more output temporally-entangled modes. Another
important advantage is a generation in a simple manner any desired
entanglement by controlling the driving field intensity.

\begin{figure}[b]
\rotatebox{0}{\includegraphics* [scale = 0.5]{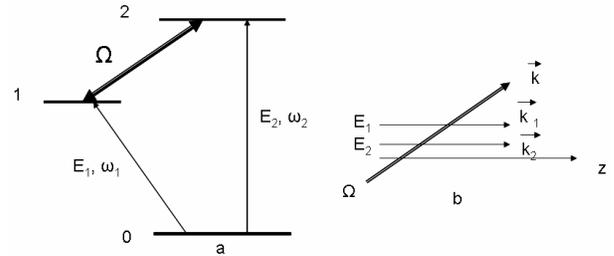}}%
\caption{(a) Level scheme of atoms interacting with quantum fields
$E_{1,2}$ and classical rf driving field of Rabi frequency $\Omega
.$ (b) Geometry of fields propagation.}
\end{figure}

We consider an ensemble of $\Delta $-type cold atoms with level
configuration as in Fig.1. Two quantum fields $$E_{1,2}^{(+)}(z,t)=\sqrt{%
\frac{\hbar \omega _{i}}{2{\varepsilon }_{0}V}}\hat{\mathcal E}%
_{1,2}(z,t)\exp [i(k_{1,2}z-\omega _{1,2}t)]$$  co-propagate along
the $z$ axis and interact with the atoms on the transitions $\mid
0$ $\rangle
\rightarrow \mid 1$ $\rangle $ \ and $\mid 0$ $\rangle \rightarrow \mid 2$ $%
\rangle ,$\ respectively, while the electric-dipole forbidden transition $%
\mid 1$ $\rangle \rightarrow \mid 2$ $\rangle $ is driven by a
classical and constant radio-frequency field (rf)with Rabi
frequency $\Omega $ inducing a magnetic dipole or an electric
quadruple transition\ between the two upper
levels. Here the electric fields are expressed in terms of the operators $\hat{%
\mathcal E}_{i}(z,t)$ obeying the commutation relations
\begin{equation}
\lbrack \hat{\mathcal E}_{i}(z,t),\hat{\mathcal
E}_{j}^{+}(z^{^{\prime }},t)]=L\delta _{ij}\delta (z-z^{^{\prime
}})
\end{equation}%
where $L$ is the length of the medium. We describe the latter using atomic
operators $\hat{\sigma}_{\alpha \beta }(z,t)$ $=\frac{1}{N_{z}}\underset{i=1}%
{\overset{N_{z}}{\sum }}\mid \alpha $ $\rangle _{i}\langle \beta
\mid $ averaged over the volume containing many atoms
$N_{z}=\frac{N}{L}dz\gg 1$ around position $z$, where $N$ is the
total number of the atoms. In the rotating wave picture the
interaction Hamiltonian is given by
$$H=-\hbar \frac{N}{L}\int\limits_{0}^{L}dz[g_{1}\hat{\mathcal E}_{1}\hat{%
\sigma}_{10}e^{ik_{1}z}+g_{2}\hat{\mathcal E}_{2}\hat{\sigma}%
_{20}e^{ik_{2}z}+\Omega \hat{\sigma}_{21}e^{ik_{\parallel }z}$$
\begin{equation}
+h.c.]
\end{equation}%
 Here $k_{\parallel }=\vec{k}_{d}\hat{e}_{z}$ \ is the
projection of the wave-vector of the driving field on the $z$
axis, $g_{\alpha }=\mu _{\alpha o}\sqrt{\frac{\hbar \omega
_{i}}{2{\varepsilon }_{0}V}}$ is the atom-field coupling constants
with $\mu _{\alpha \beta }$ being the dipole matrix
element on the transition $\mid \alpha $ $\rangle \rightarrow \mid \beta $ $%
\rangle $ , and $V$ is the quantization volume taken to be equal
to interaction volume. For simplicity, we discuss the case of
exactly resonant interaction with all fields and, therefore, put
in Eq.(1) the frequency detunings equal to zero, neglecting so the
Doppler broadening, which in a cold atomic sample is smaller than
all relaxation rates. Then, using the slowly varying envelope
approximation, the propagation equations for the quantum field
operators take the form:

\begin{equation}
\left( \frac{\partial }{\partial z}+\frac{1}{c}\frac{\partial }{\partial t}%
\right) \hat{\mathcal E}_{1}(z,t)=ig_{1}\frac{N}{c}\hat{\sigma}%
_{01}e^{-ik_{1}z}+\hat{F}_{1}
\end{equation}%
\
\begin{equation}
\left( \frac{\partial }{\partial z}+\frac{1}{c}\frac{\partial }{\partial t}%
\right) \hat{\mathcal E}_{2}(z,t)=ig_{2}\frac{N}{c}\hat{\sigma}%
_{02}e^{-ik_{2}z}+\hat{F}_{2}
\end{equation}%
where $\hat{F}_{i}(z,t)$ are the commutator preserving Langevin operators,
whose explicit form is not of interest here.

In the weak-field (single-photon) limit, the equation of atomic coherences $%
\hat{\rho}_{0i}=\hat{\sigma}_{0i}e^{-ik_{i}z},\ \ i=1,2~\ \ \ $and$\ \ \ \ \
\hat{\rho}_{12}=\hat{\sigma}_{12}e^{-i(k_{2}-k_{1})z}\ $ are treated
perturbatively in $\hat{\mathcal E}_{1,2}$. In first order only $\hat{\sigma%
}_{00}\simeq 1$ is different from zero and for these equations we
get:
\begin{equation}
\frac{\partial }{\partial t}\hat{\rho}_{01}=-\Gamma _{1}\hat{\rho}%
_{01}+ig_{1}\hat{\mathcal E}_{1}\hat{\sigma}_{00}+i\Omega ^{\ast }\hat{\rho}%
_{02}e^{i\bigtriangleup kz}-ig_{2}\hat{\mathcal
E}_{2}\hat{\rho}_{21}
\end{equation}%
\begin{equation}
\frac{\partial }{\partial t}\hat{\rho}_{02}=-\Gamma _{2}\hat{\rho}%
_{02}+ig_{2}\hat{\mathcal E}_{2}\hat{\sigma}_{00}+i\Omega ^{\ast }\hat{\rho}%
_{01}e^{-i\bigtriangleup kz}-ig_{1}\hat{\mathcal
E}_{1}\hat{\rho}_{12}
\end{equation}%
\begin{equation}
\frac{\partial }{\partial t}\hat{\rho}_{12}=-(\Gamma _{1}+\Gamma _{2})\hat{%
\rho}_{12}-ig_{1}\hat{\mathcal E}_{1}^{\ast }\hat{\rho}_{02}+ig_{2}\hat{%
\mathcal E}_{2}\hat{\rho}_{10}
\end{equation}%
Here $2\Gamma _{1,2}$ are the decay rates of the excited states $\mid 1$ $%
\rangle $ and $\mid 2$ $\rangle $ and $\Delta k=k_{2}-k_{1}-k_{\parallel }$
is the wave-vector mismatch.

Further, we assume that the phase-matching condition $\Delta k=0$
is fulfilled in the medium. Then, the solution to Eqs.(5-7) to the
first order in $\hat{\mathcal E}_{1,2}$ is readily found to be
\begin{equation}
\hat{\rho}_{01}=-i\frac{\Gamma }{D}g_{1}\hat{\mathcal
E}_{1}-i\frac{\Omega
^{2}-\Gamma ^{2}}{D^{2}}g_{1}\frac{\partial }{\partial t}\hat{\mathcal E}%
_{1}-\frac{\Omega }{D}g_{2}\hat{\mathcal E}_{2}+\frac{2\Gamma \Omega }{D^{2}%
}g_{2}\frac{\partial }{\partial t}\hat{\mathcal E}_{2},
\end{equation}%
\begin{equation}
\hat{\rho}_{02}=\hat{\rho}_{01}(1\leftrightarrow 2),\ \ \ \text{ }D=\Omega
^{2}+\Gamma ^{2}.
\end{equation}%
where, for simplicity, the optical decay rates are taken to be the
same: $\Gamma _{1}=\Gamma _{2}=\Gamma $. The first terms in right
hand side (RHS) of Eqs.(8,9) are responsible for linear absorption
of quantum fields and define the field absorption coefficients
$k_{i}=\frac{g_{i}^{2}\Gamma N}{c\Omega ^{2}}$ upon substituting
these expressions into Eqs.(3,4). Here the condition of
electromagnetically induced transparency (EIT,
refs.\cite{harris,mfleis}) $\Omega \gg \Gamma _{1,2}$ is assumed
to be satisfied for both transitions with weak-field coupling. The
second terms in RHS of Eqs.(8,9) represent the dispersion
contribution to the group velocities of the pulses, while the two
rest terms describe the parametric interaction between the fields.
We require the
photon absorption to be strongly reduced by imposing the condition $%
k_{i}L\ll 1.$ Another limitation follows from $\Delta \omega _{EIT}T\geq 1$
indicating that the initial spectrum of quantum fields is contained within
the EIT window $\Delta \omega _{EIT}=\frac{\Omega ^{2}}{\Gamma }$ $\frac{1}{%
\sqrt{\alpha }}$ \cite{fleis}, where $T$ is a duration of weak-field pulses,
$\alpha =\mathcal{N}\sigma L$ is optical depth, $\sigma =\frac{3}{4\pi }%
\lambda ^{2}$ is resonant absorption cross-section and $\mathcal{N}$ is the
atomic number density. Finally, the length of the pulses has to fit the
length of the medium: $Tv_{i}<L$ with $v_{i}=\frac{c\Omega ^{2}}{g_{i}^{2}N}$
being the group velocity of the $i$-th field. Taking into account that $%
k_{i}L\sim \Gamma ^{2}\alpha /\Omega ^{2},$ this set of limitations yields
\begin{equation}
\frac{\Omega ^{2}}{\Gamma ^{2}}>>\alpha \text{ \ \ \ \ and \ }\frac{1}{\sqrt{%
\alpha }}\ll \frac{Tv_{i}}{_{L}}<1\text{\ \ \ }
\end{equation}%
\ It is worth noting that upon satisfying the conditions (10), the
dominant contribution to the parametric coupling between the
photons is the third term in RHS of Eq.(8,9), because in this case
the last term becomes strongly suppressed by the factor $\Omega
^{2}T/\Gamma >>1.$

It is useful at this point to consider numerical estimations. The
sample is chosen to be $^{85}Rb$ vapor with the ground state
$5S_{1/2}(F_{g}=3)$ and exited states $5P_{3/2}(F_{e}=2),$
$5P_{3/2}(F_{e}=3)$ of $Rb$ atom as the states $\mid 0$ $\rangle $
and $\mid 1$ $\rangle ,\mid 2$ $\rangle $ \ in Fig.1 ,
respectively, using the following parameters: light wavelength
$\lambda
\simeq 0.8\mu $m$,$ $\Gamma =2\pi \ast 3$ MHz, atomic density $\mathcal{N}%
\sim 10^{12}$cm$^{-3}$ in a trap$\ $of length $L\sim 100$ $\mu $m,
$\Omega \sim 10\Gamma ,$ and the input pulse duration $T\simeq $
2$\div $3ns. In this
case $\alpha \simeq 16,$ $v_{2}\sim 10^{4}m/s$, $v_{1}\sim 0.3v_{2},$ and $%
k_{i}L\leqslant 0.2.$ All of the parameters we use in our
calculations appear to be within experimental reach, including
initial single-photon wave packets with a duration of several ns
satisfying the narrow-line limitation discussed above. The
standard method for producing single photons based on SPDC in
nonlinear crystals does not fit our purpose due to too broad
linewidth ($\sim 10$nm) of generated light. Recently, a source of
narrow-bandwidth, frequency tunable single photons with properties
allowing to excite the narrow atomic resonances has been created
\cite{chou, eis}.

Then, taking into account that in the absence of photon losses the
noise operators $\hat{F}_{i}$ in Eqs.(5) give no contribution, the
simple propagation equations for the field operators are finally
obtained:
\begin{equation}
\left( \frac{\partial }{\partial z}+\frac{1}{v_{1}}\frac{\partial
}{\partial t}\right) \hat{\mathcal E}_{1}(z,t)=-i\beta
\hat{\mathcal E}_{2}
\end{equation}%
\begin{equation}
\left( \frac{\partial }{\partial z}+\frac{1}{v_{2}}\frac{\partial
}{\partial t}\right) \hat{\mathcal E}_{2}(z,t)=-i\beta
\hat{\mathcal E}_{1}
\end{equation}%
where $\beta ={g_{1}g_{2}N/c\Omega }$ is the parametric coupling
constant. It is easy to check that these equations preserve the
commutation relations (1). Note that for the parameters above, the
parametric interaction between the photons is highly strong $\beta
L\sim 3$.

The formal solution of Eqs.(11,12) for field operators in the region $%
0\leqslant z\leqslant L$ is written as
\begin{equation*}
\hat{\mathcal E}_{i}(z,t)=\hat{\mathcal E}_{i}(0,\
t-z/v_{i})+\int\limits_{0}^{z}dx\{\hat{\mathcal E}_{i}(0,\ t-z/v_{i}-\frac{%
\Delta v_{ij}}{v_{i}v_{j}}(z-x))
\end{equation*}%
\begin{equation}
\ast \frac{\partial J_{0}(\psi )}{\partial z}-i\beta \ \hat{\mathcal E}%
_{j}(0,\ t-z/v_{i}-\frac{\Delta v_{ij}}{v_{i}v_{j}}(z-x))\
J_{0}(\psi )\},\
\end{equation}%
where $i,j=1,2$ \ and $j\neq i.$ The Bessel function $\ J_{0}(\psi
)$ depends on $z$ via $\psi =2\beta \sqrt{x(z-x)}$ , $\Delta
v_{ij}=v_{i}-v_{j}$ is the difference of group velocities.

We are interested in dynamics of input state $\mid \psi _{in}$ $\rangle =$ $%
\mid 1_{1}$ $\rangle \otimes \mid 0_{2}$ $\rangle $ containing one
photon in $\omega _{1}$ field. The similar results are clearly
obtained in the case of one input photon at $\omega _{2}$
frequency. We assume that initially the
$\omega _{1}$ pulse is located around $z=0$ with a given temporal profile $%
f_{1}(t):$%
\begin{equation}
\langle 0\mid \hat{\mathcal E}_{1}(0,t)\mid \psi _{in}\rangle
=\langle 0\mid \hat{\mathcal E}_{1}(0,t)\mid 1_{1}\rangle
=f_{1}(t)
\end{equation}%
The intensities of quantum fields at any time are given by%
\begin{equation}
\langle I_{i}(z,t)\rangle =\mid \langle 0\mid \hat{\mathcal
E}_{i}(z,t)\mid \psi _{in}\rangle \mid ^{2}
\end{equation}%
Using Eqs.(13-15) and recalling that $\langle 0\mid \hat{\mathcal E}%
_{2}(0,t)\mid \psi _{in}\rangle =0$, we calculate $\langle
I_{i}\rangle $ numerically and show in Fig.2 the output pulse
shapes at $\ z=L$ for the three values of $\Omega $ and for the
case of Gaussian input $($at $z=0)$ pulse $f_{1}(t)=\exp
[-2t^{2}/T^{2}]$ .
\begin{figure}[b]
\rotatebox{0}{\includegraphics* [scale = 1]{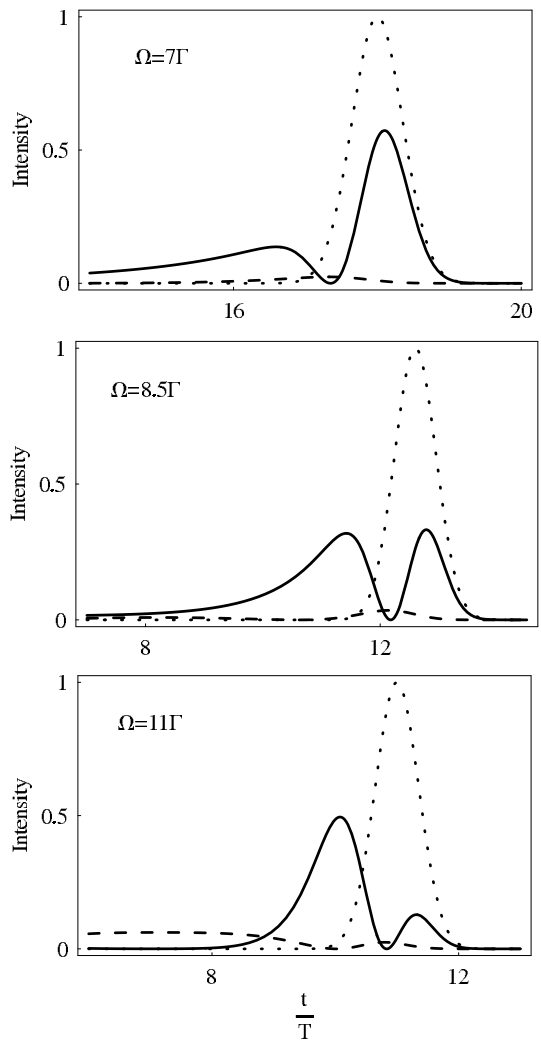}}%
\caption{The numerical solution to Eqs.(15) at the output of the
medium $z=L$ for three values of $\ \Omega $. In these figures,
solid curves
represent the results for $\omega _{1}$ pulse, dashed curves show the $%
\omega _{2}$ field generated in the medium, and dotted lines
correspond to initial Gaussian pulse at $\omega _{1}$ frequency
with $T=2$ns propagating in the medium in the absence of
parametric interaction $\beta =0.$ For the rest of parameters see
the text.}
\end{figure}
For one-photon initial state, as is the case here, one can clearly
see that the second\ field is not practically generated, thus
demonstrating that our scheme enables to prepare
a single-photon in a pure temporally-delocalized state with an efficiency $%
\sim 100\%.$ Moreover, depending on the driving field intensity, a
different degree of initial pulse splitting and, hence, of
entanglement is attainable. It is evident also that the total
number of photons which is determined by the areas of the
corresponding peaks is conserved upon propagation through the
medium. Besides, in this case only two well-separated output
temporal modes at $\omega _{1}$ frequency are produced, where due to $%
v_{2}>v_{1}$ a newly generated component is advanced compared to
the signal pulse. This separation depends on the relative velocity
of quantum fields, the larger the ratio $v_{2}/v_{1}$ the larger
the group delay and the larger the output pulses separation. On
the contrary, in the limit of equal group velocities the
propagating pulses experience no splitting, as it can be easily
seen from Eqs.(13), which in this case are reduced to
\begin{equation}
\hat{\mathcal E}_{i}(z,t)=\hat{\mathcal E}_{i}(0,\tau)cos (\beta
z)-i\hat{\mathcal E}_{j}(0,\tau)sin (\beta z)
\end{equation}%
where $\tau=t-z/v_{1}$, $j\neq i.$

The system displays, however, a much richer dynamics in the case
of input state $\mid \psi _{in}$ $\rangle =$ $\mid 1_{1}$ $\rangle
\otimes \mid 1_{2}$ $\rangle $
consisting of one-photon wave packets at both frequencies $\omega _{1}$ and $%
\omega _{2}.$ These results will be published elsewhere. Here we
note only that in this case two multi-time-bin qubits at different
frequencies $\omega _{1}$ and $\omega _{2}$ are generated, being
at the same time strongly correlated with each other. This is
evident also from the particular result of Eq.(16).

The single-photon states are completely described by their Wigner
function, whose remarkable property is that it takes negative
values at the origin of phase space for the complex field
amplitude. The negativity of the Wigner function is the ultimate
signature of non-classical nature of these states. Besides, the
nonlocality of quantum correlations between the two temporal modes
directly follows from the violation of Bell's inequality
$-2\leqslant \mathcal B\leqslant 2 $  predicted by local theories
\cite{bonas}. Here the combination $\mathcal B$ has the form:
\begin{equation*}
\mathcal B=\frac{\pi ^{2}}{4}[W(0,0)+W(\alpha _{1},0)+W(0,\alpha
_{2})-W(\alpha _{1},\alpha _{2})]
\end{equation*}
where
\begin{equation}
W(\alpha _{1},\alpha _{2})=\frac{4}{\pi ^{2}}[2\mid \alpha
_{1}+\alpha _{2}\mid ^{2}-1] e^{-2\mid \alpha _{1}\mid ^{2}-2\mid
\alpha _{2}\mid ^{2}}
\end{equation}%
is the Wigner function of two temporal modes calculated for the
values of complex amplitudes $\alpha _{i}=x_{i}+iy_{i}$ with
$x_{i}$ and $y_{i}, i=1,2,$ being the quadratures of the $i$-th
mode. In Eq.(16) we have supposed zero relative phase between
superposition amplitudes of the two modes. In our case, for
experimental verification of Bell's theorem, the Wigner function
can be derived from the data of homodyne detection of quantum
fields, when the signals at the detectors are measured at two
different times matched to the time separation between two $\omega
_{1}$ output pulses obtained in Fig.2.

In conclusion, we have studied a highly efficient scheme for
dynamic preparation of a single photon in distinct temporal modes,
employing strong parametric interaction between two single-photon
pulses, under the conditions of EIT, and their group delay. We
have found the solution of propagation equations for the field
operators depending on the propagation distance in terms of the
Bessel function, the oscillatory character of which is just
responsible for pulse temporal splitting. We have shown the
ability of our scheme to achieve an arbitrary entanglement by
adjusting the driving field intensity, while the separation
between the time bins can be controlled by using the different
atomic-level configurations to obtain the different ratio of group
velocities of quantum fields. Subsequent papers will discuss the
more complicated case of two input single-photon pulses and will
present the results of detail numerical simulations.

This work was supported by the ISTC Grant No.A-1095 and, in part,
by Research Grant 0047 of the Armenian Republic Government.

\end{document}